# Inkjet printed circuits with two-dimensional semiconductor inks for high-performance electronics

Tian Carey[1,2*†], Adrees Arbab[1,2†] Luca Anzi[3†], Helen Bristow[2], Fei Hui[4], Sivasambu Bohm[1,2], Gwenhivir Wyatt-Moon[5], Andrew Flewitt[5], Andrew Wadsworth[2], Nicola Gasparini[2], Jong M. Kim[5], Mario Lanza[6], Iain McCulloch[7,8], Roman Sordan[3*], Felice Torrisi[1,2*]

[1]Cambridge Graphene Centre, Department of Engineering, University of Cambridge, UK

[2]Department of Chemistry, Molecular Sciences Research Hub, Imperial College London, White City Campus, Wood Lane, London W12 0BZ, UK

[3]L-NESS, Department of Physics, Politecnico di Milano, Via Anzani 42, 22100 Como, Italy

[4]Department of Materials Science and Engineering, Technion - Israel Institute of Technology, Haifa 3200003, Israel

[5]CAPE Building, Department of Engineering, University of Cambridge, UK

[6] Physical Sciences and Engineering Division, King Abdullah University of Science and Technology (KAUST), Thuwal 23955-6900, Saudi Arabia

[7]KAUST Solar Center (KSC), King Abdullah University of Science and Technology (KAUST), Thuwal, 23955-6900 Saudi Arabia

[8]Department of Chemistry, Chemistry Research Laboratory, University of Oxford, Oxford, OX1 3TA, UK

† T. C., A.A. and L.A. have contributed equally to this work.
* Correspondence and requests for materials should be addressed to
 F.T. (f.torrisi@imperial.ac.uk), R.S. (roman.sordan@polimi.it), T.C. (tian.carey@cantab.net)

ABSTRACT

Air-stable semiconducting inks suitable for complementary logic are key to create low-power printed integrated circuits (ICs). High-performance printable electronic inks with two-dimensional materials have the potential to enable the next generation of high performance, low-cost printed digital electronics. Here we demonstrate air-stable, low voltage (< 5 V) operation of inkjet-printed n-type molybdenum disulfide ($MoS_2$) and p-type indacenodithiophene-*co*-benzothiadiazole (IDT-BT) field-effect transistors (FETs), estimating a switching time of $\tau_{MoS2}$ ~ 3.3 µs for the $MoS_2$ FETs. We achieve this by engineering high-quality $MoS_2$ and air-stable IDT-BT inks suitable for inkjet-printing complementary pairs of n-type $MoS_2$ and p-type IDT-BT FETs. We then integrate $MoS_2$ and IDT-BT FETs to realise inkjet-printed complementary logic inverters with a voltage gain $|A_v|$ ~ 4 when in resistive load configuration and $|A_v|$ ~ 1.36 in complementary configuration. These results represent a key enabling step towards ubiquitous long-term stable, low-cost printed digital ICs.

INTRODUCTION

Digital ICs mainly rely on a metal-oxide-semiconductor (MOS) technology that use p-type (PMOS logic) or n-type (NMOS logic) field-effect transistors (FETs) to implement mixed signal ICs[1] and logic gates.[2] A complementary MOS (CMOS) logic uses both p-type and n-type FETs and enables ICs with drastically smaller power dissipation than that of PMOS and NMOS logic.[3] Roughly 95% of all modern ICs use CMOS logic,[4] making it a fundamental technology towards low-power, scalable circuits.[4] Figures of merit such as field effect mobility ($\mu$), on/off current ratio ($I_{on}/I_{off}$), switching time ($\tau$) and inverter voltage gain ($|A_v|$), defined as the slope of the inverter voltage transfer characteristic ($dV_{OUT}/dV_{IN}$, where $V_{IN}$ is the input and $V_{OUT}$ the output voltage), have been used to assess and benchmark the performance of the FETs and ICs, respectively. Ultra-large-scale integration circuits such as memories, microcontrollers and large-area flexible active-matrix displays rely on CMOS and resistive load (i.e. PMOS and NMOS) logic circuits requiring $|A_v| > 1$, which are achievable in ambient conditions without the use of high-vacuum ($< 10^{-6}$ mbar) technologies.[5]

Printing electronic circuits have the potential to enable low-cost ubiquitous IC systems on arbitrary substrates for flexible[6] and wearable electronics.[7] The key for creating high-performance printed FETs is to ideally obtain a highly crystalline semiconductor channel,[8,9] improve printing uniformity (i.e. decrease the active channel roughness)[9] and remove residual surfactants and solvents from the printed films.[10] Furthermore, optimal printed CMOS logic circuits with $|A_v| > 1$ require reducing the FET threshold voltage, $V_{th} < 1$ V to reduce the power supply voltage and the same drain current $I_D$ levels for both p-type and n-type FETs.[11] Printable electronic inks achieving such performances with conducting, semiconducting and insulating or dielectric properties is key to satisfy the diverse nature IC components, such as interconnects, resistors, diodes, capacitors and transistors.[12,13] Inkjet printing is a prime technique for large-area fabrication of printed electronics, combining advantages such as versatility to print on wide range of substrates (e.g. textile, polymers and silicon), mask-less and non-contact fabrication with a high resolution (~ 50 μm), manufacturing scalability (m$^2$/minute) and low material losses (< 1 ml)[7], which make it a well-established technique to print ICs based on PMOS, NMOS and CMOS technologies.[14]

The majority of scientific efforts towards inkjet printed FETs and CMOS logic is focused on the development of organic polymers, metal oxides and sorted single walled carbon nanotubes (SWCNTs) as the channel material.[15–17] For example, poly(3-hexylthiophene) (P3HT) has been used as a p-type channel material with an n-type material poly{[N,N9-bis(2-octyldodecyl)-naphthalene-1,4,5,8-bis(dicarboximide)-2,6-diyl]-alt-5,59-(2,29-bithiophene)} (P(NDI2OD-T2)) to enable an organic polymer inkjet printed CMOS inverter on polyethylene terephthalate (PET). The FETs had $\mu$ < 1 cm$^2$V$^{-1}$s$^{-1}$, $I_{on}/I_{off}$ ~ 10$^6$ and $|A_v|$ ~ 25.[18] However, organic polymers frequently suffer from long-term (> 100 h) stability issues in ambient condition[19] and n-type organic polymers have struggled to exceed $\mu$ ~ 1 cm$^2$V$^{-1}$s$^{-1}$ after decades of research.[20] Doped SWCNTs have been used to achieve inkjet printed FETs with $\mu$ ~ 2 cm$^2$V$^{-1}$s$^{-1}$, $I_{on}/I_{off}$ ~ 10$^5$ and CMOS inverters with $|A_v|$ ~ 85 on a Si/SiO$_2$ substrate.[16] Inkjet printed doped metal oxides have also been investigated. Inkjet printed zinc tin oxide was used as an n-type material ($\mu$ ~ 4 cm$^2$ V$^{-1}$ s$^{-1}$, $I_{on}/I_{off}$ ~ 10$^6$) with inkjet printed p-type SWCNT ($\mu$ ~ 2 cm$^2$ V$^{-1}$ s$^{-1}$, $I_{on}/I_{off}$ ~ 10$^4$) to enable a CMOS inverter with $|A_v|$ ~ 17.1.[17] Unfortunately, SWCNTs require semiconducting vs metallic sorting before deposition, limiting manufacturability.[16]

Electronic few-layer two-dimensional (2D) material inks (E2D inks) offer properties suitable to inkjet print ICs.[21] Dispersions of 2D materials can be mass-produced by liquid-phase exfoliation (LPE) and formulated into electronic few-layer 2D material inks E2D inks[21], with conducting (e.g. MXenes), semiconducting (e.g. transition metal dichalcogenides) and insulating properties (e.g. hexagonal boron nitride, h-BN or silicates) suitable for printed and conformable electronics on-demand and in scale[22]. For example, semiconducting molybdenum disulfide ($MoS_2$) or tungsten diselenide ($WSe_2$) inks can be used as the active layer of transistors[23] or photodetectors.[24] E2D inks report excellent chemical stability in ambient conditions,[25] tuneable p-type and n-type semiconductors[26] and scalable production methods[27] paving the way to the next generation of solution processed electronics. Inkjet printed graphene FETs reached $\mu$ of up to 95 cm$^2$ V$^{-1}$ s$^{-1}$ and $I_{on}/I_{off}$ ~10, on a surface-modified Si/SiO$_2$ substrate.[10] Fully inkjet-printed dielectrically-gated flexible graphene/h-BN FETs reported µ of up to 204 cm$^2$ V$^{-1}$ s$^{-1}$ on polyethylene terephthalate (PET) with $I_{on}/I_{off}$ ~2.5 at low operating voltage (< 5 V) in ambient conditions.[10] These graphene/h-BN FETs enabled inkjet-printed ICs such as memories, logic gates and CMOS with a gain of only $|A_v|$ ~ 0.1.[10] Current modulation in printed thin films of LPE $MoS_2$, tungsten disulfide ($WS_2$), molybdenum diselenide ($MoSe_2$) and $WSe_2$ on PET has been attempted via electrochemical gating by a liquid electrolyte (LE), achieving $I_{on}/I_{off}$ ~ 600 and $\mu$ ~ 0.1 cm$^2$ V$^{-1}$s$^{-1}$ for $MoSe_2$ and $WS_2$ films in vacuum[28]. However, the absence of printable highly crystalline semiconducting E2D inks able to exhibit field-effect modulation in ambient conditions has impeded the implementation of inkjet printed 2D material FETs suitable for digital ICs.[21,29]

Recently, improved solution-processing resulted in the production of high-quality semiconducting 2D materials by electrochemical intercalation of bulk $MoS_2$ layered crystals via quaternary ammonium molecules in acetonitrile.[30] Spin coated thin-films of the $MoS_2$ achieved $\mu$ of up to 10 cm$^2$ V$^{-1}$s$^{-1}$ and $I_{on}/I_{off}$ ~ 10$^6$ in vacuum with NMOS circuits with $|A_v|$ ~ 20. However, the low concentration (< 0.2 mg/ml) $MoS_2$ ink in dimethylformamide (DMF) makes them unsuitable for any scalable printing (such as inkjet). Inkjet printed CMOS logic has not yet been achieved with semiconducting E2D inks. In this work, we demonstrate circuits of CMOS and NMOS logic using inkjet printed n-type $MoS_2$ and p-type IDT-BT FETs achieving mobility of ~ 0.06 cm$^2$ V$^{-1}$ s$^{-1}$ at low voltage (< 5 V) and fast switching times

($\tau_{MoS2}$ ~ 3.3 μs). The printed logics achieved $|A_v|$ ~ 4 and $|A_v|$ ~ 1.36 in NMOS and CMOS configurations, respectively.

**RESULTS AND DISCUSSION**

**Inks formulation**

We use inkjet printing as a versatile method to manufacture logic circuits. Ink properties such as surface tension $\gamma$, viscosity $\eta$, boiling point $B_p$ and concentration $c$ must be engineered to ensure satellite droplet free ejection, high throughput and morphologically uniform (i.e. roughness minimisation) printed films.[7,10] The jetting of ink from an inkjet printer is determined by $\eta$, $\gamma$, density $\rho$ and the nozzle diameter, $a$.[31] The inverse of the Ohnesorge number ($Oh$) has been traditionally used as a figure of merit $Z = Oh^{-1} = (\gamma \rho a)^{1/2}/\eta$ to characterize the quality of droplet formation from the nozzle of an inkjet printer.[7] An optimal $Z$ range to minimize the formation of satellite droplets (i.e. secondary droplets that are produced in addition to the first droplet ejected from the inkjet nozzle)[31] from an inkjet nozzle has been identified as $2 < Z < 24$. Therefore, we will engineer our ink to be within the optimal $Z$ range to avoid any jetting issues. Inks with $c$ ~ 1 mg/ml are required to maximise throughput while $\gamma$ ~ 30 mN/m is optimal to coalesce the droplets after deposition. $B_p < 100°C$ ensures a quick evaporation of the solvent minimising both the transport of particulates (which causes 'coffee ring') and the re-dispersion of material after multiple passes, thus improving the morphological uniformity of the printed film.[10,32]

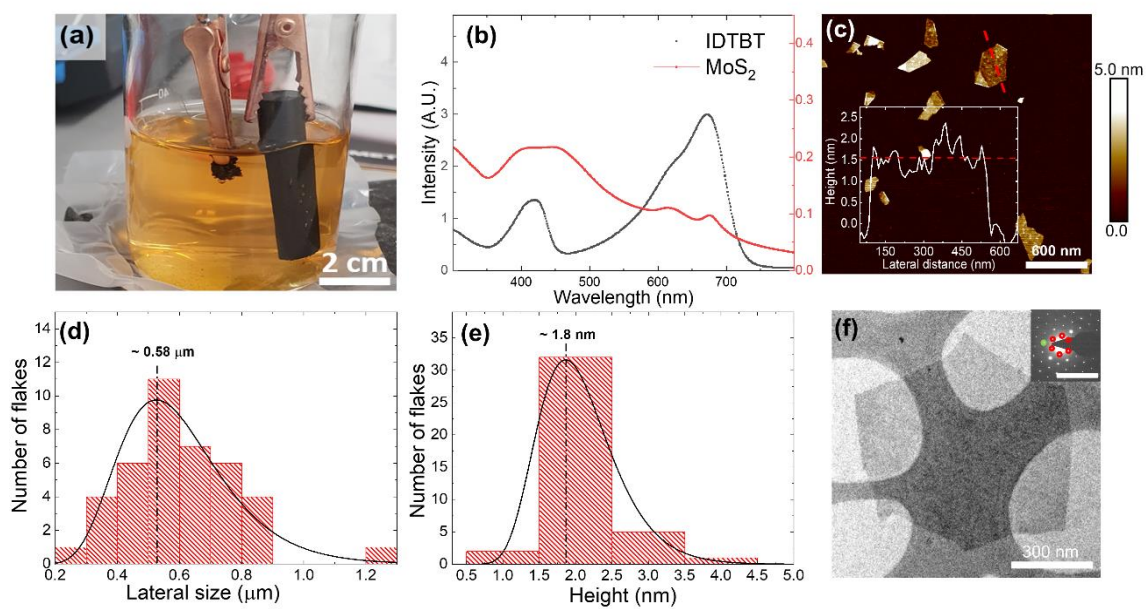

Figure 1 (a) Electrochemical setup used to exfoliate an MoS2 crystal. (b) UV-Vis spectra of the MoS2 and IDT-BT. (c) AFM micrograph of MoS2 flakes with the scale indicating the maximum height being no more than 5 nm for the majority of flakes. The cross section of one of the MoS2 flakes (red dashed line) is shown in the inset having a height of ~ 1.5 nm. Lateral (d) and height (e) distribution of the MoS2 flakes dispersed in IPA with the black curve showing the lognormal distribution of each dimension. (f). TEM image of an MoS2 flake along with the selective area diffraction pattern on the top right hand corner with the inset scale bar representing 10 nm. The diffraction pattern shows the (100) plane (red circles) and the (110) plane (green circle) which highlights the six-fold-symmetry.

We intercalate the $MoS_2$ crystal with tetraheptylammonium (TA) cations to separate single and few-layer $MoS_2$ from the bulk crystal in acetonitrile/quaternary ammonium bromide electrolyte (figure 1a).[30] The few-layer $MoS_2$ is then ultrasonicated in DMF with polyvinylpyrrolidone (PVP) (see methods). The $MoS_2$ is then solvent exchanged into isopropanol/PVP (IPA/PVP) by decanting the DMF and pipetting IPA/PVP into the sedimented $MoS_2$. We do this to engineer the inkjet printable n-type $MoS_2$ ink with $\eta_{MoS2}$ ~ 1.8 mPa s, $\gamma_{MoS2}$ ~ 28 mN m$^{-1}$, $\rho_{MoS2}$ ~ 0.7 g cm$^{-3}$, consistent with our previous reports.[10,22] We select as a p-type ink the copolymer IDT-BT, synthesised according to the previously reported results[33] to use as a p-type material. We choose IDT-BT as it has previously demonstrated stability in air ambient for up to 100h.[33] For printing the IDT-BT, we optimise the ink by dispersing at $c_{\text{ID-TBT}}$ ~ 6 mg/ml in 1,2-dichlorobenzene (DCB) with $\eta_{\text{ID-TBT}}$ ~ 16 mPa s, $\gamma_{\text{IDTBT}}$ ~ 20.6 mN m$^{-1}$, $\rho_{\text{ID-TBT}}$ ~ 1.3 g cm$^{-3}$. We find $Z$ ~ 11 and $Z$ ~ 2 for the MoS$_2$ and IDT-BT inks, respectively, which fall within the optimal printing range.[10,22]

Figure 1b shows the UV-visible optical absorption spectra of the MoS$_2$ and IDT-BT inks. We use the Beer-Lambert law to correlate the absorbance (red line) to the $c$ of MoS$_2$ flakes (see methods). We find $c$ ~ 2.65 mg/ml for the MoS$_2$ ink (using an absorption coefficient $\alpha_{\text{MoS2}}$ ~ 3400 L g$^{-1}$ m$^{-1}$).[34] The spectra of the MoS$_2$ flakes displays two characteristic peaks at 610 nm and 672 nm attributed to the A exciton and B exciton, respectively, confirming the presence of MoS$_2$ flakes and consistent with the previous reports.[35] UV-vis spectroscopy also confirms the presence of IDT-BT, figure 1b (black line) shows the characteristic peaks which are associated with the IDT-BT copolymer. The most prominent peak appears at 672 nm and is attributed the backbone of the C$_{16}$-IDT-BT polymer.[33,36]

The lateral size ($<\underline{L}>$) and thickness ($t$) of the MoS$_2$ flakes is estimated using atomic force microscopy (AFM) statistics. An AFM micrograph of several MoS$_2$ flakes deposited on Si/SiO$_2$ substrate is shown in figure 1c. The cross section of one of the MoS$_2$ flakes (red dashed line) is shown in the inset of figure 1c with a thickness of 1.5 nm and lateral size of ~ 550 nm. Figures 1d, e show the log-normal distribution [37,38] of $<\underline{L}>$ and $t$ over 40 MoS$_2$ flakes. The $t$ peaks at 1.8 nm (corresponding to an average number of layers, $N$ ~ 3)[39] and $<\underline{L}>$ peaks at 0.58 µm. A typical transmission electron microscopy (TEM) is shown in figure 1f and a selected area electron diffraction image shown in the inset. The TEM agrees with the AFM data showing a lateral size of ~ 1 µm while the diffraction pattern indicates the six-fold symmetry that is expected from MoS$_2$ flakes.[22] The MoS$_2$ $<\underline{L}>$ determined from both TEM and AFM is about 2% the size of the nozzle diameter (21 μm), hence matching with the requirements of drop-on-demand (DOD) the inkjet printing[40].

**Inkjet printed MoS₂ FETs**

Before printing, we use electron-beam (e-beam) lithography to pattern ~ 1-mm wide source, drain and gate electrodes on a Si/SiO$_2$ substrate (see methods). We use Ti/Au (5/35 nm) for the source and drain and Al (40 nm) for gate electrodes of the FETs. A thin native AlO$_x$ layer (oxide thickness $t_{ox}$ ~ 4 nm) is formed at the top surface of Al by air exposure, creating an Al/AlO$_x$ gate stack.[41] The gate length $L$ ~ 500 nm, with the source-to-drain distance ~ 100 nm longer. We use a DOD inkjet printer to print the semiconductor channel and define the MoS$_2$ and IDT-BT FETs (figure 2a) as follows. The MoS$_2$ ink is printed for 30 printing passes (i.e. where one pass is defined as an area with ink droplets deposited 30 μm from each other) over the source, drain and gate electrodes to create the MoS$_2$ channel of the FET. We use an interdrop spacing (i.e. the centre-to-centre distance between two adjacent deposited droplets) of 30 μm with a platen temperature of 20 °C. We print a channel width, $W$ = 400 μm. Profilometry measurements (see methods), reveal a channel thickness ($t_c$) as ~ 20 nm, which matches the percolation threshold of inkjet printed E2D inks.[22] We use scanning electron microscopy (SEM) and AFM to confirm the MoS$_2$ flakes are covering the source, gate and drain electrodes (figure 2a, inset). Further investigation with conductive AFM (see methods) indicates that only a portion of the MoS$_2$ flakes (from 5.56% to 13.01% depending on the portion of channel scanned, on average ~9.25%) produce effective charge transport between the source and the drain electrodes, leading to an estimated effective $W$ ~ 40 μm.

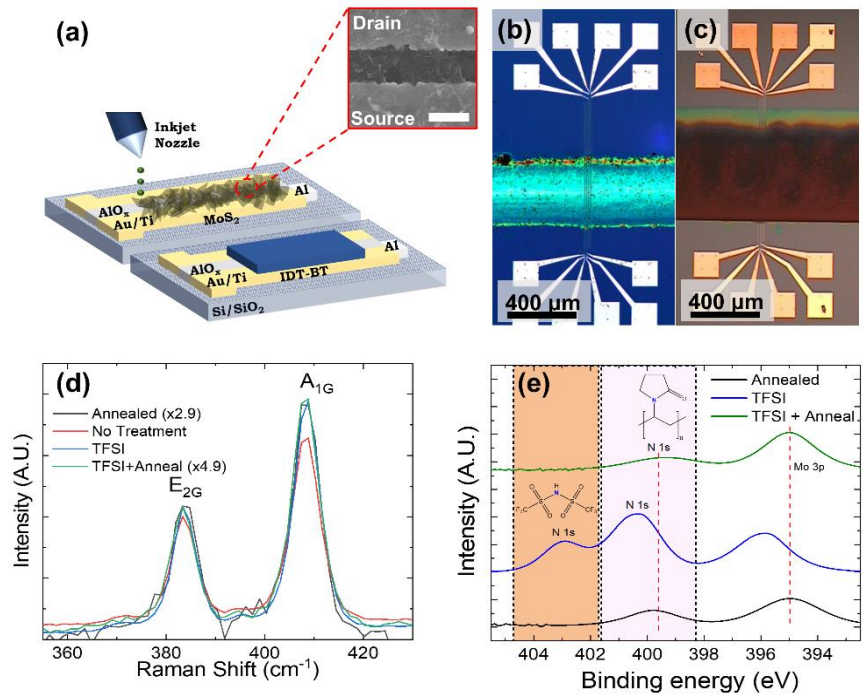

Figure 2 (a) Schematic of the FETs with MOS$_2$ and IDT-BT inkjet printed channels across Ti/Au and Al/AlO$_x$ contacts. SEM image of the printed MoS$_2$ channel which shows MoS$_2$ flakes bridging the Ti/Au and Al/AlO$_x$ contacts, scale bar is 200 nm (inset). Optical microscope image of the (b) MoS$_2$ and (c) IDT-BT devices. (d) Raman spectra of the inkjet printed MoS$_2$ channel at different treatment conditions, showing the characteristic E$_{2g}$ and A$_{1g}$ peaks. (e) XPS spectra of MoS$_2$ flakes after different treatment conditions.

A post-processing treatment is required for the MoS$_2$ FETs to remove PVP stabilisation agent after inkjet printing. We submerge our devices in bis(trifluoromethane)sulfonimide (TFSI) in 1,2-dichloroethane at 100°C for 1 hour. The treatment is performed in a nitrogen glovebox to avoid exposing TFSI to moisture in ambient atmosphere.[42] The devices are then annealed at 400 °C for 1 hour to remove any residual solvent (see methods). Thermo-gravimetric analysis is undertaken on the PVP stabilisation agent (see methods) which indicates PVP starts to thermally decompose at ~350 °C, with the rate of decomposition peaking at 434 °C indicating that the annealing step could also be assisting in the removal of PVP from the MoS$_2$ channel. We inkjet printed the IDT-BT ink at an interdrop spacing of 30 μm and platen temperature of 20°C with 20 printing passes resulting in the same $t_c$ (~300 nm) as in the previous work.[33] The $W$ = 600 μm for IDT-BT FETs. Optical microscopy is used to examine the inkjet printed MoS$_2$ (figure 2b) and IDT-BT (figure 2c) FETs. We observe that the films are printed over the source, drain and gate electrodes as expected.

Raman spectroscopy is used to monitor the quality of the inkjet printed MoS$_2$ channel following the individual or combined effect of TFSI and annealing treatment. Figure 2d plots the spectra of the MoS$_2$ film as-deposited (red curve), after annealing and no TFSI treatment (black curve), after TFSI treatment and no annealing (blue curve) and after TFSI treatment followed by annealing (green curve). All spectra show the typical MoS$_2$ A$_{1g}$ and E$_{2g}$ peaks at 406 cm$^{-1}$ and 383 cm$^{-1}$, respectively.[43] The frequency difference (23 cm$^{-1}$) between A$_{1g}$ and E$_{2g}$ confirms an average thickness of few layers (3-4 layers).[43] The position of the E$_{2g}$ peak, Pos(E$_{2g}$) ~ 383 cm$^{-1}$, indicates the absence of a large number of defects at every step of our post-processing treatment.[44]

X-ray photoelectron spectroscopy (XPS) is also used to determine the effect that TFSI treatment and annealing have on the MoS$_2$ inkjet printed films. Figure 2e shows the XPS spectra of the MoS$_2$ ink in the energy region of 393 - 405 eV when TFSI treated (blue curve), annealed (black curve) and combined TFSI treatment followed by annealing (green line). We select this region of the XPS spectra as it allows to examine the respective changes in the binding energy of MoS$_2$ electrons for the three samples. The annealed MoS$_2$ sample (black curve) shows a peak at 395 eV, which corresponds to the Mo 3p peak and is associated with molybdenum in MoS$_2$.[45] The position of the Mo 3p peak coincides in the TFSI treated and annealed MoS$_2$ sample (green curve), which upshifts to 396 eV in the TFSI treated MoS$_2$ samples (blue curve). We attribute this to the electronegative fluorine atom present in the molecular structure of TFSI. The fluorine atom attracts electron density towards itself from other surrounding atoms in molecules present in the thin film, hence, reducing the shielding of the nuclear charge and causing the binding energy of nearby atoms to increase.[46] Besides the Mo 3p peak, the three curves exhibit an N 1s orbital peak at ~ 400 eV attributed to the nitrogen atom present in PVP[47] and the blue curve shows a second N 1s peak at 403 eV attributed to nitrogen atoms in TFSI.[48,49] Interestingly, while these peaks are quite intense in the blue curve, they are heavily suppressed in the green curve, indicating the removal of TFSI and a significant reduction of PVP concentration (with respect to the black curve), which also shows the same peak but significantly reduced. This indicates that the combination of TFSI and annealing treatments is necessary to minimise the presence of PVP which can otherwise degrade the electrical performance of our inkjet printed $MoS_2$ FET.

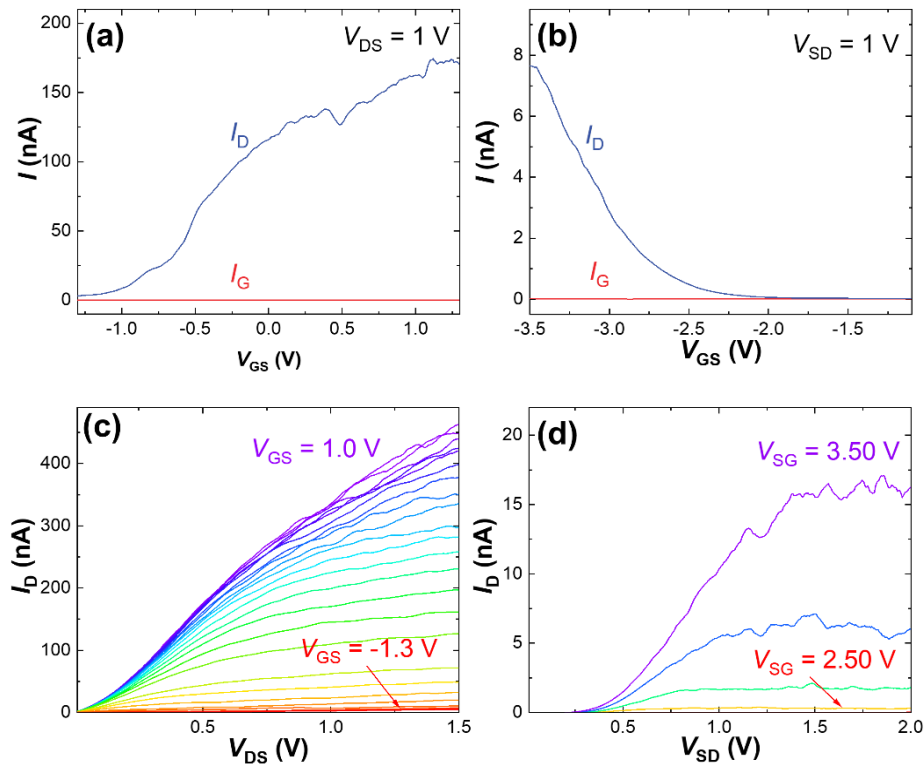

Figure 3 The transfer curve of the (a) $MoS_2$ and (b) IDT-BT FET which plots $I_D$ as a function of $V_{GS}$ with $|V_{DS}| = 1$ V. The gate current $I_G$ is simultaneously measured (red curve). (c) The output curves of the $MoS_2$ FET where the drain current $I_D$ is measured as a function of $V_{DS}$ (at $V_{GS}$ ranging from -1.3 V to 1 V with a step change of 0.1 V). (d) The output curves of the IDT-BT FET with $I_D$ measured as a function of $V_{SD}$ (at $V_{SG}$ ranging from 2.5 V to 3.5 V with a step change of 0.25 V).

We characterise the electrical properties of the inkjet printed $MoS_2$ FETs and IDT-BT FETs using a probe station in air (see methods). For both FETs, the transfer characteristics (the drain current, $I_D$ as a function of the gate-source voltage $V_{GS}$) (figure 3a, b) are measured applying a drain-source voltage $|V_{DS}| = 1$ V (note that fluctuations in the curve are typical of all low-current devices). The $MoS_2$ FETs exhibit an n-type and IDT-BT FETs p-type behaviour. The output characteristic ($I_D$ as a function of the drain-source voltage, $V_{DS}$) of the $MoS_2$ FET (figure 3c) is measured at $V_{GS}$ ranging from -1.3 V to 1.0 V with a step of 0.1 V. Conversely, the output curve of the IDT-BT FET (figure 3d) is measured at $V_{SG}$ ranging from 2.5 V to 3.5 V with a step of 0.25 V. These curves are consistent with the behaviour of $MoS_2$ FETs[30] and IDT-BT FETs[33] prepared by spin coating on $Si/SiO_2$ and glass substrates, respectively. The gate current in all the devices is $I_G < 100$ pA, which is three orders of magnitude smaller than the drain current (> 100 nA), indicating that the device is modulating current in the semiconductor channel. Additionally, in the transfer curves (figure 3a, b), the forward and backward sweep indicate minimal hysteresis. The field-effect µ of the $MoS_2$ and IDT-BT FETs were derived from equation $\mu = (L/W)(1/C)(g_m/V_{DS})$, where $g_m = \partial I_D/\partial V_{GS}$ is the transconductance. We estimate the dielectric capacitance per unit area $C \sim 95$ nFcm$^{-2}$ as an effective medium approximation of $AlO_x$ and air attributed to the variation of the peak to valley ($R_{pp} \sim 33$ nm) roughness of the gate for the $MoS_2$ FETs. We assume that the IDT-BT conforms to the gate roughness and is only gated by $AlO_x$. The ambient field-effect mobility for $MoS_2$ and IDT-BT FETs are calculated to be up to $\mu_{MoS2} \sim 0.06$ cm$^2$ V$^{-1}$ s$^{-1}$ and $\mu_{IDT-BT} \sim 2\times10^{-4}$ cm$^2$ V$^{-1}$ s$^{-1}$, with $I_{on}/I_{off} \sim 50$ and $10^3$ respectively. The measured $\mu_{MoS2}$ is the extrinsic field effect mobility of the inkjet printed $MoS_2$ FET and we believe the value could be higher, if the contact resistance were to be reduced. The $\mu$ is comparable to that of inkjet printed $MoS_2$ electrochemical transistors ($\mu \sim 0.1$ cm$^2$ V$^{-1}$ s$^{-1}$)[23] operating under high vacuum and lower than spin coated IDT-BT ($\mu \sim 1$ cm$^2$ V$^{-1}$ s$^{-1}$)[33] which we attribute to the higher roughness ($R_{pp} \sim 80$ nm) of the IDT-BT channel.

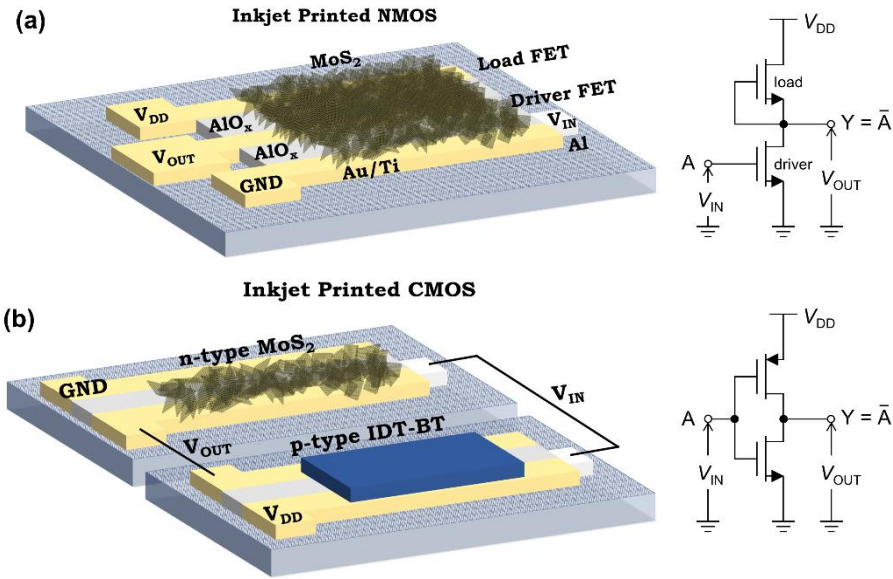

Figure 4 Schematic and circuit diagram of the inkjet printed (a) NMOS and (b) CMOS inverters.

The versatility of inkjet-printed n-type and p-type FETs allows to easily design printed electronic circuits. Printed circuits with semiconducting E2D inks have been a long-sought milestone on the way towards the next generation of printed electronics. To verify the suitability of our technology to print circuits with 2D materials and the versatility to mutually operate with organic electronics, we demonstrate inkjet-printed NMOS and CMOS logic implementing the MoS$_2$ FETs in n-type MoS$_2$ depletion-load inverters (Fig.4a) and MoS$_2$/IDT-BT CMOS inverters (Fig.4b) shown in the schematic.

Figure 5a shows the static voltage transfer characteristic (the output voltage $V_{OUT}$ as a function of the input voltage $V_{IN}$, blue line) and the corresponding low-frequency voltage gain $A_{v,NMOS}$ (red line, the scale on the right) of the NMOS inverter at the power supply voltage $V_{DD} = 1.5$ V. The device correctly operates as an inverter with a highest voltage gain

$|A_{v,NMOS}| \sim 4$. Figure 5b shows the digital waveforms measured in the NMOS inverter. The input voltage swing is equal to $V_{DD}$ (as in conventional logic gates) while the output voltage swing is ~ 73 % $V_{DD}$. The mismatch between $V_{IN}$ and $V_{OUT}$ logic voltage levels is attributed to the negative threshold voltage of the MoS$_2$ FETs.

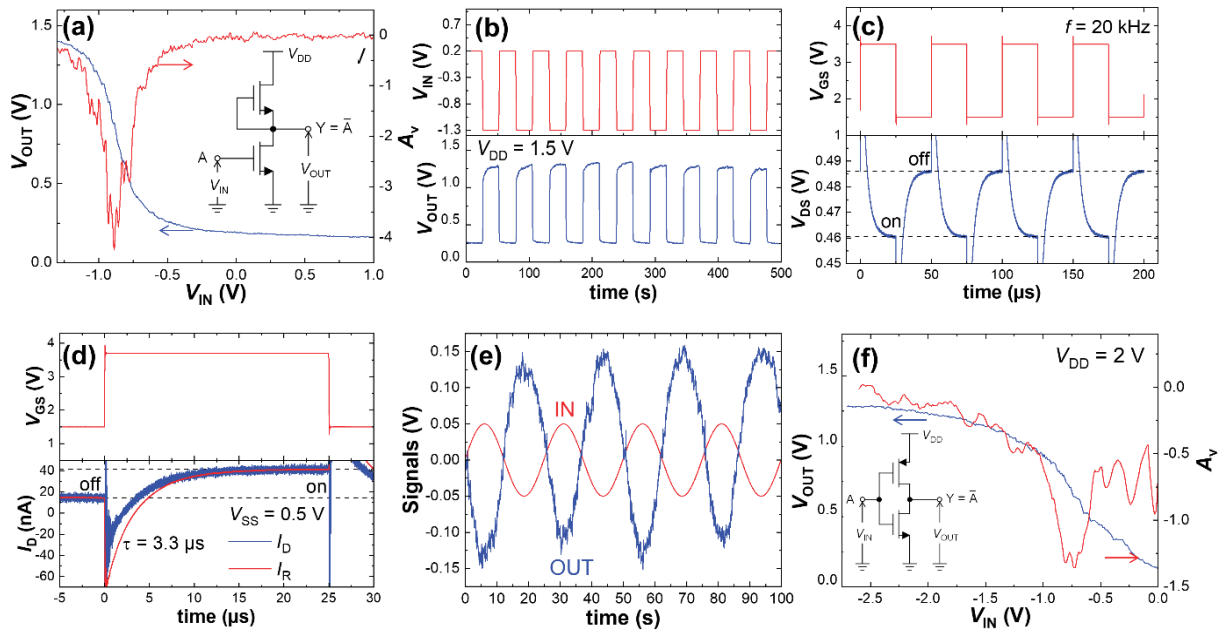

Figure 5 (a) The static voltage transfer characteristic and the corresponding low-frequency frequency voltage gain ($A_v$) of an NMOS inverter. The circuit diagram shows the NMOS inverter made of a driver MoS$_2$ FET (bottom) and a load MoS$_2$ FET top. (b) The digital $V_{IN}$ (red) and $V_{OUT}$ (blue) waveforms measured in the NMOS inverter as a function of time at $V_{DD} = 1.5$ V. (c) The $V_{DS}$ waveforms measured in the circuit comprising a single MoS$_2$ FET. $V_{GS}$ is a square wave at 20 kHz. The spikes of $V_{DS}$ are a consequence of the capacitive currents in the circuit. (d) The same measurement in high temporal resolution reveals $\tau_{MoS2} \sim 3.3$ µs from the exponential decay fit. The plot shows both the drain current $I_D$ and the resistive component of the drain current $I_R$ (which is proportional to the drain potential). The difference between $I_D$ and $I_R$ is the capacitive component of the drain current $I_C$ which approaches to zero in the steady state. (e) The waveforms measured in the NMOS inverter operating as a small-signal voltage amplifier. The offsets are removed for clarity. (f) The transfer characteristic of the fabricated CMOS inverter and the corresponding voltage gain at $V_{DD} = 2$ V.

The switching time ($\tau_{NMOS}$), defined as the *RC* time constant of the output voltage in figure 5b, is $\tau_{NMOS} \sim 400$ ms which is dominated by the large channel resistance (~ 200 MΩ) and capacitive load of the BNC cables used to connect the FETs to the probe station. In contrast to FETs with liquid-electrolyte gates,[23] the fabricated gates are capable of operating at high-frequency[50] and therefore do not significantly contribute to the time constant of the $MoS_2$ FETs. Figure 5c shows $V_{DS}$ of an individual $MoS_2$ FET driven by a square wave $V_{GS}$ at a frequency of 20 kHz. The time constant of a single FET is mostly dominated by the drain time constant, i.e. by the parasitic drain capacitances of the FET and drain load. This leads to the much smaller time constant $\tau_{MoS2}$ ~ 3.3 µs (figure 5d) compared to the NMOS inverter (due to the off-chip connections used to connect the FETs in the circuit). The $\tau_{MoS2}$ here is smaller than that of the only printed $MoS_2$ transistors ($\tau$ ~ 68 ms) reported on literature.[23] However, this is expected given the latter is an electrochemical transistor where the switching time is limited by the mass transfer of ions, unlike our FETs.

We also operate the inverter as a small-signal voltage amplifier as shown by the analogue waveforms measured in figure 5e. The inverter is biased at the operating point where it exhibits the highest voltage gain (i.e., at $V_{IN}$ ~ -0.9 V). The amplitude of the output voltage (blue curve) is ~ 4 times larger than the amplitude of the input sine wave voltage (red curve), i.e., $|A_{v,NMOS}|$ ~ 4.

We also realised CMOS inverters made of the inkjet printed IDT-BT p-type FETs and $MoS_2$ n-type FETs with a power supply voltage $V_{DD} = 2$ V (figure 5f). Figure 5f shows the transfer characteristics of the CMOS inverter and the corresponding low-frequency voltage gain $A_{v,CMOS}$ of the CMOS inverter at the power supply voltages $V_{DD}$ of 2 V. Also in this case the device correctly operates as an inverter, giving rise to $|A_v|_{CMOS} > 1$ at $V_{DD} = 2$ V. We find a maximum $A_{v,CMOS}$ ~ -1.4 at $V_{DD} = 2$ V, which is higher than inkjet printed logic circuits previously reported with graphene inks, as expected.[10]

**CONCLUSION**

We demonstrated air-stable and low voltage (< 5 V) inkjet printed n-type MoS$_2$ with a fast switching time of $\tau_{MoS2}$ ~ 3.3 µs, which is four orders of magnitude faster than state-of-the art solution processed MoS$_2$ transistors.[23]. We integrated the MoS$_2$ FETs in NMOS logic and, with p-type IDT-BT FETs, in CMOS logic circuit architectures. The CMOS and NMOS inverters demonstrated a voltage gain of $|A_v|_{,CMOS}$ ~ 1.4 and $|A_v|_{,NMOS}$ ~ 4 , respectively, which is ~ 40 times greater than that of inkjet printed inverters with graphene inks.[10] The assembly of stable printed complementary logic combining 2D materials with organic polymers is a fundamental building block towards ubiquitous, low-cost and high-performance digital electronics, which is manufacturable in large scale.

## METHODS

**Electrochemical exfoliation of MoS$_2$ and ink formulation:** An electrochemical cell with two electrodes is used to intercalate a MoS$_2$ crystal (HQ graphene). Copper crocodile clips are used to mount the anode and cathode electrode. A thin piece of MoS$_2$ crystal is used as the cathode while a graphite rod (Qingdao *Tennry* Carbon Co.) is used as the anode. Tetraheptylammonium bromide (Sigma-Aldrich, CAS number: 4368-51-8, SKU: 87301) 5 mg/ml is added to acetonitrile (~ 50 ml) which acted as the electrolyte. The MoS$_2$ crystal and graphite rod should be completely submerged in the electrolyte. We apply a voltage of 8V and allow the MoS$_2$ to be intercalated for 1 hour. Further details on the reaction mechanism can be found in Lin et al.[51]

After the reaction, the MoS$_2$ was intercalated with tetraheptylammonium cations and expanded increasing the volume of material on the crocodile clip. The MoS$_2$ was then washed with ethanol and manually broken into smaller pieces. The MoS$_2$ was then sonicated in DMF with PVP (Sigma-Aldrich, CAS Number 9003-39-8, SKU: PVP40) 22 mg/ml (Molecular weight ~ 40,000) for 30 minutes. To further assist the break-up of material, a shear mixer (IKA-T10) was then used for 10 minutes at 10,000 rpm. The dispersion was then centrifuged at 3,000 rpm for 20 minutes (g ~ 1534, r-max = 15.25 cm) to remove large chunks (> 10 layers) of $MoS_2$ material and then a further 5,000 rpm for 10 min (g ~ 4193, r-max = 15.25 cm). Both centrifuge steps made use of a ProteomeLab™ XL-A by Beckman Coulter coupled with an SW 32 Ti Swing-Bucket rotor. The resulting dispersion is a dark green colour indicating the successful exfoliation of a few-layer $MoS_2$. Attempts to disperse the material directly in ethanol were not successful as the initial centrifugation step resulted in complete sedimentation. Therefore, the larger chunks could not be removed.

The $MoS_2$/PVP/DMF dispersion was placed under high vacuum (operating at ~ 1 mbar) at 40 °C for 240 minutes to concentrate 500 ml into 10 ml. The concentrated dispersion was then centrifuged at 45,000 rpm for 30 min (g ~ 216886, r-max = 9.58 cm) in a Beckman Optima MAX Ultracentrifuge to sediment the $MoS_2$. The $MoS_2$ sediment was then redispersed in IPA with PVP (22 mg/ml (Molecular weight ~ 40,000) to create the $MoS_2$ ink.

**Formulation and coating of IDT-BT:** The hexadecyl alkyl side chain substituted IDT-BT was synthesised according to previously reported literature.[33] The IDT-BT polymer was dissolved in 1,2 dichlorobenzene at a $c$ ~ 6 mgml$^{-1}$ for inkjet printing.

**Fabrication of electrodes and dielectric on a Silicon Wafer**: Highly doped Si chips with a thermally grown 90-nm-thick $SiO_2$ top layer were used in the fabrication of the $MoS_2$ FETs. The chips were initially cleaned by ultrasonication in acetone/isopropanol. The organic contamination from the surface of the chips was removed by $O_2$ plasma in a Tepla 300 AL PC plasma asher. Devices were patterned by e-beam lithography using a fixed-beam moving-stage (FBMS) mode of a Raith eLINE system at 30 keV. The FBMS mode allowed to pattern very wide (950 µm) and very short (65 nm) electrodes without any stitching errors. Source and drain contacts (with a contact length of 1 µm and width $W$ = 950 µm) were first defined by e-beam lithography followed by e-beam evaporation of Ti/Au (5/35 nm) in an e-beam evaporator at a base pressure of ~ $1.2 \times 10^{-6}$ mbar. Local back-gates (with a gate length $L$ of 600 nm) were patterned by e-beam lithography in the second step followed by e-beam evaporation of Al (40 nm) in the e-beam evaporator at a base pressure of ~ $1.2 \times 10^{-6}$ mbar. A thin (~4 nm) native $AlO_x$ layer was formed at the top surface of Al after the samples were kept in air ambient for one day.

**Inkjet Printing**: A drop-on-demand inkjet printer (Fujifilm Dimatix DMP-2800) using a 21 µm diameter nozzle (Fujifilm DMC-11610) was used to print devices. For both $MoS_2$ and IDT-BT inks, one nozzle was used for printing with a droplet volume of ~10 pL. An interdrop spacing of 30 µm with a platen temperature of 20 °C is used. Thirty printing passes (i.e. where one pass is defined as an area with ink droplets deposited 30 µm from each other) are used create the $MoS_2$ channels while twenty printing passes are used for the IDT-BT ink. The reduced number of printing passes is chosen so $I_D$ in the on state are matching in $MoS_2$ and IDT-BT FETs. Both inks were printed with a maximum jetting frequency of 2 kHz.

**Film treatment**: The $MoS_2$ transistors were treated with 10 mg/ml of TFSI (Sigma-Aldrich, CAS Number: 82113-65-3, SKU: 15220) in 1,2-dichloroethane (Acros Organics) at 100 °C for 1 hour.[52] This is done in a nitrogen glovebox to minimise the exposure of TFSI to air. The film is then heated on a hot plate (in $N_2$ atmosphere) for 1 hour at 400 °C.

**Atomic Force Microscopy:** A Bruker Dimension Icon working was used to analyse the area and thickness of the $MoS_2$ flakes contained in the E2D ink. The $MoS_2$ ink was drop cast onto a $SiO_2$ substrate after dilution by a factor of 1:100. The samples were scanned in peakforce mode, and 40 $MoS_2$ flakes were counted to determine the statistics for the lateral size and thickness. The lateral size is calculated as the square root of the flake length times the flake width. A NX-HighVac AFM working in tapping-contact mode (provided with a NCHR tip from NanoWorld) was used to analyse the surface roughness of the $MoS_2$ channel. The same apparatus was dotted with a solid Pt tip from Rocky Mountain (RMN-25PT300B) to collect current maps of the devices in contact mode, by applying a positive voltage to the AFM tip and keeping the drain electrode grounded.

**Scanning electron microscopy**: The scanning electron microscope (SEM) imaging was performed in Raith eLINE at 10 kV.

**Profilometry:** A Bruker DektakXT stylus profilometer was used to get the step height of the 30 layer printed films on $Si/SiO_2$ with a stylus tip radius of 12.5 µm and stylus force of 3 mg. The resolution of the profile was 0.666 µm/point and speed 200 µm/s. The data was obtained using Vision64.ink software.

**Probe station Measurement**: All electrical measurements were performed in air ambient in FormFactor probe stations EP6 and Summit 11000. The electrical characterizations of the FETs and inverters were performed by Keithley 2600 series source-measure units at a typical sweep rate of 145 mV/s. The time constant was measured using a Tektronix AFG 3022B arbitrary function generator, Keysight DSO9404A digital storage oscilloscope (bandwidth 4 GHz), and Keysight N2795A active probe (bandwidth 1 GHz).

**Raman Spectroscopy:** Films of $MoS_2$ ~200 nm thick are deposited onto $Si/SiO_2$ substrate, and a Raman spectrum is obtained using a Renishaw 1000 InVia micro-Raman spectrometer at 514.5 nm using a ×50 objective and incident power ~1 mW.

**Optical absorption spectroscopy**: The flake concentration of the $MoS_2$ ink is found using the Beer-Lambert law which correlates the absorbance $A = \alpha c l$, to the flake concentration $c$, the absorption coefficient $\alpha$ and the light path length $l$. The $MoS_2$ ink is diluted at a 1:200 ratio with DMF/PVP. An absorption coefficient of $\alpha_{MoS_2}$ ~ 3400 L g$^{-1}$ m$^{-1}$ for the MoS$_2$ ink at 660 nm is used.[53]

**Surface Tension:** The pendant drop method (First Ten Angstroms FTA1000B) is used to measure the MoS2 surface tension. Drop shape analysis is used to fit a shadow image of a droplet suspended from a needle and the surface tension is calculated using the Young–Laplace equation.[32]

**Rheometry:** A parallel plate rotational rheometer (DHR rheometer TA instruments) is used to calculate the infinite-rate viscosity of the MoS$_2$ ink. The measurement was taken at 25 °C.

**Optical Microscope:** An optical microscope (Nikon Optiphot 300 attached to Schott ACE 1 light source) is used to image deposited droplets in dark field mode. The images are acquired at a ×20 magnification.

**Transmission Electron Microscopy:** The TEM was performed with a FEI Philips Tecnai F-20 operated at 200 kV (Tungsten, LB6) with a line resolution of 0.10 and point resolution of 0.24 nm. Spherical aberration Cs (objective): 1.2 mm. The IPA/PVP-MoS$_2$ ink was diluted by a factor of 100 and drop cast on holey carbon film 400 mesh copper (Cu) grid (Agar scientific product code AGS147-4) and allowed to dry naturally overnight.

**X-ray photoelectron spectroscopy:** K-Alpha+ surface analysis by Thermo scientific was used to obtain the XPS data of the MoS2 films. The spectra were obtained under $8 \times 10^{-7}$ mbar (ultra-vacuum) which employs micro-focused Al Kα X-ray source (1486 eV) and a 2D detector attached to 180 double-focusing hemispherical analyser. The data obtained were further analysed using Avantage software by Thermo scientific.

**Thermogravimetric Analysis:** TGA measurements of the PVP were performed with a TA Instruments Q50 under nitrogen with a ramping temperature of 10 °C min$^{-1}$ from ~ 25 °C up

to 1000 °C.

**Acknowledgements:**

The authors acknowledge funding from EPSRC grants EP/P02534X/2, EP/R511547/1, EP/T005106/1, Imperial College Collaboration Kick-Starter grant and Trinity College, Cambridge, the EU H2020 Graphene Flagship Core 3 Grant No. 881603, and a Technion-Guangdong Fellowship.